\documentclass[review,12pt,authoryear]{elsarticle}

\usepackage{amssymb}
\usepackage{amsthm}
\usepackage{amsmath,color}
\usepackage{mathrsfs}
\usepackage{graphicx}
\usepackage{epstopdf}
\usepackage{float}
\usepackage{caption}
\usepackage{subcaption}
\usepackage{bm}
\usepackage{bbm}
\usepackage{mathrsfs}
\usepackage{cleveref}
\usepackage{soul}
\usepackage{accents}
\usepackage{color,soul} 
\usepackage{color} 
\soulregister\citep7 
\soulregister\citet7 
\soulregister\citealp7 

\journal{}

\makeatletter
\def\@author#1{\g@addto@macro\elsauthors{\normalsize%
    \def\baselinestretch{1}%
    \upshape\authorsep#1\unskip\textsuperscript{%
      \ifx\@fnmark\@empty\else\unskip\sep\@fnmark\let\sep=,\fi
      \ifx\@corref\@empty\else\unskip\sep\@corref\let\sep=,\fi
      }%
    \def\authorsep{\unskip,\space}%
    \global\let\@fnmark\@empty
    \global\let\@corref\@empty  
    \global\let\sep\@empty}%
    \@eadauthor={#1}
}
\makeatother

\begin{document}

\begin{frontmatter}



\title{On fracture in finite strain gradient plasticity}


\author{E. Mart\'{\i}nez-Pa\~neda\corref{cor1}\fnref{Uniovi}}
\ead{martinezemilio@uniovi.es}

\author{C. F. Niordson\fnref{DTU}}

\address[Uniovi]{Department of Construction and Manufacturing Engineering, University of Oviedo, Gij\'on 33203, Spain}

\address[DTU]{Department of Mechanical Engineering, Solid Mechanics, Technical University of Denmark, DK-2800 Kgs. Lyngby, Denmark}

\cortext[cor1]{Corresponding author. Tel: +34 985 18 19 67; fax: +34 985 18 24 33.}

\begin{abstract}
In this work a general framework for damage and fracture assessment including the effect of strain gradients is provided. Both mechanism-based and phenomenological strain gradient plasticity (SGP) theories are implemented numerically using finite deformation theory and crack tip fields are investigated. Differences and similarities between the two approaches within continuum SGP modeling are highlighted and discussed. Local strain hardening promoted by geometrically necessary dislocations (GNDs) in the vicinity of the crack leads to much higher stresses, relative to classical plasticity predictions. These differences increase significantly when large strains are taken into account, as a consequence of the contribution of strain gradients to the work hardening of the material. The magnitude of stress elevation at the crack tip and the distance ahead of the crack where GNDs significantly alter the stress distributions are quantified. The SGP dominated zone extends over meaningful physical lengths that could embrace the critical distance of several damage mechanisms, being particularly relevant for hydrogen assisted cracking models. A major role of a certain length parameter is observed in the multiple parameter version of the phenomenological SGP theory. Since this also dominates the mechanics of indentation testing, results suggest that length parameters characteristic of mode I fracture should be inferred from nanoindentation.
\end{abstract}


\end{frontmatter}


\section{Introduction}
\label{Introduction}

Experiments and direct dislocation simulations have shown that metallic materials display strong size effects at the micron and sub-micron scales. Attributed to geometrically necessary dislocations (GNDs) associated with non-uniform plastic deformation, this size effect is especially significant in fracture problems as the plastic zone adjacent to the crack tip may be physically small and contains large spatial gradients of deformation.\\

Much research has been devoted to modeling experimentally observed size effects (e.g., \citealp{FH93,NH03,B10,K13}) and several continuum strain gradient plasticity (SGP) theories have been proposed through the years in order to incorporate length scale parameters in the constitutive equations. Of particular interest from the crack tip characterization perspective is the development of formulations within the finite deformation framework (e.g., \citealp{GA05,G08,P09}). In spite of the numerical complexities associated, various studies of size effects under large strains have been conducted using both crystal \citep{KT08,B14} and isotropic \citep{NR04,L07,MR09,A12} gradient-enhanced plasticity theories. Isotropic SGP formulations can be classified according to different criteria, one distinguishing between phenomenological theories \citep{FH97,FH01} and microstructurally/mechanism-based ones \citep{G99,Q03}.\\

The experimental observation of cleavage fracture in the presence of significant plastic flow \citep{E94,K02} has encouraged significant interest in the role of the plastic strain gradient in fracture and damage assessment. Studies conducted in the framework of phenomenological \citep{WH97,K08,N12} and mechanism-based theories \citep{WX05,S07} have shown that GNDs near the crack tip promote local strain hardening and lead to a much higher stress level as compared with classical plasticity predictions. However, although large deformations take place in the vicinity of the crack, the aforementioned studies were conducted within the infinitesimal deformation theory and little work has been done to investigate crack tip fields modeled by SGP accounting for finite strains. \citet{H03} developed a finite deformation framework for the mechanism-based strain gradient (MSG) plasticity theory but were unable to reach strain levels higher than 10\% near the crack tip due to convergence problems. \citet{PY11} used the element-free Galerkin method to characterize crack tip fields through a lower order gradient plasticity (LGP) model \citep{YC00}. From a phenomenological perspective, \citet{TN08} analyzed the influence of the strain gradient at a crack tip interacting with a number of voids while \citet{MG09} determined the range of material length scales where a full strain gradient dependent plasticity simulation is necessary.\\
 
Very recently, \citet{MB15} identified and quantified the relation between material parameters and the physical length over which gradient effects prominently enhance crack tip stresses from a mechanism-based approach. The numerical results obtained in \citet{MB15} show a significant increase in the differences between the stress fields of MSG and conventional plasticity when finite strains are taken into account. This is due to the strain gradient contribution to the work hardening of the material, which lowers crack tip blunting and thereby suppresses the local stress triaxiality reduction characteristic of conventional plasticity predictions \citep{M77}. These results revealed the important influence of strain gradients on a wide range of fracture problems, being particularly relevant in hydrogen assisted cracking modeling due to the central role that the stress field close to the crack tip plays on both hydrogen diffusion and interface decohesion. Moreover, Gangloff and his co-workers have shown that accurate correlations with experimental measurements can be achieved by adopting high levels of hydrostatic stress from dislocation-based micromechanical modeling of hydrogen embrittlement \citep{T03,LG07,G14}.\\ 

In this paper crack tip fields are evaluated thoroughly with both phenomenological and mechanism-based strain gradient plasticity theories with the aim of gaining insight into the role of the increased dislocation density associated with large gradients in plastic strain near the crack. Differences between the two main classes of SGP theories are examined and their physical implications discussed. In both approaches the numerical scheme is developed to allow for large strains and rotations providing an appropriate framework for damage and fracture assessment within SGP theories.

\section{Material models}
\label{Phenomenological and MSG theories}

The key elements of the two SGP theories considered in this work are summarized in this section, with particular emphasis on the constitutive equations and other aspects of interest from the fracture mechanics perspective. Comprehensive details, including the variational formulation and the corresponding differential equations, can be found in \citep{FH01, NH03} and \citep{G99,Q03} for the phenomenological and mechanism-based cases, respectively.\\

\subsection{Fleck and Hutchinson's gradient theory}
\label{Fleck and Hutchinson's gradient theory}

The strain gradient generalization of $J_2$ flow theory proposed by \citet{FH01} is considered to model size effects in metal plasticity from a phenomenological perspective. In this theory hardening effects due to plastic strain gradients are included through the gradient of the plastic strain rate $\dot{\varepsilon}^p_{ij,k}=\left(m_{ij} \dot{\varepsilon}^p \right)_{,k}$. Where $\dot{\varepsilon}^p=\sqrt{\frac{2}{3} \dot{\varepsilon}_{ij}^p \dot{\varepsilon}_{ij}^p}$ is the increment in the conventional measure of the effective plastic strain and $m_{ij}=\frac{3}{2}s_{ij}/\sigma_e$ is the direction of the plastic strain increment, with $s_{ij}$ denoting the stress deviator, and $\sigma_e$ the von Mises effective stress. The gradient enhanced effective plastic strain rate, $\dot{E}^p$ can be defined in terms of three unique, non-negative invariants of $\dot{\varepsilon}^p_{ij,k}$, which are homogeneous of degree two:

\begin{equation}
\dot{E}_p=\sqrt{ \dot{\varepsilon}^{p^2} + l_1^2 I_1 + l_2^2 I_2+l_3^2 I_3}
\end{equation}

\noindent where, $l_1$, $l_2$ and $l_3$ are material length parameters. The effective plastic strain rate can be expressed explicitly in terms of $\dot{\varepsilon}^p$ and $\dot{\varepsilon}_{,i}^p$:

\begin{equation}
\dot{E}_p=\sqrt{\dot{\varepsilon}^{p^2} + A_{ij} \dot{\varepsilon}_{,i}^p \dot{\varepsilon}_{,j}^p+ B_i \dot{\varepsilon}_{,i}^p \dot{\varepsilon}^p+C \dot{\varepsilon}^{p^2}}
\end{equation}

\noindent where the coefficients $A_{ij}$, $B_i$ and $C$ depend on the three material length parameters as well as on the spatial gradients of the plastic strain increment direction (for details see \citealp{FH01}).

By the alternative definitions $A_{ij}=l^{*^2}$, $B_i=0$ and $C=0$ a single length scale parameter theory closely related to the strain gradient theory of \citet{A84} can be formulated using a new length parameter $l^*$ with

\begin{equation}
\dot{E}_p=\sqrt{\dot{\varepsilon}^{p^2} + l^{*^2} \dot{\varepsilon}_{,i}^p \dot{\varepsilon}_{,i}^p}
\end{equation}

For a body of volume $V$ and surface $S$, with outward normal $n_i$, the principle of virtual work in the current configuration is given by

\begin{equation}\label{PVW}
\int_V \left(\sigma_{ij} \delta \dot{\varepsilon}_{ij} - \left(Q - \sigma_e \right) \delta \dot{\varepsilon}^p + \zeta_i \delta \dot{\varepsilon}^p_{,i} \right) \textnormal{dV} = \int_S \left( T_i \delta \dot{u}_i + t \delta \dot{\varepsilon}^p \right)\textnormal{dS}
\end{equation}

Here $\dot{u}_i$ is the displacement rate, $\dot{\varepsilon}_{ij}$ is the strain rate, $\sigma_{ij}$ denotes the Cauchy stress tensor, $Q$ is a generalized effective stress (work conjugate to the plastic strains) and $\zeta_i$ is the higher order stress (work conjugate to the plastic strain gradients). The surface integral contains traction contributions from the conventional surface traction $T_i=\sigma_{ij} n_j$ and the higher order traction $t=\zeta_i n_i$.

\subsection{Mechanism-based strain gradient (MSG) plasticity}
\label{Gao et al. theory}

The theory of mechanism-based strain gradient plasticity \citep{G99,Q03} is based on the Taylor dislocation model \citep{T38} and therefore the shear flow stress $\tau$ is formulated in terms of the dislocation density $\rho$ as

\begin{equation}\label{Eq1MSG}
\tau = \alpha \mu b \sqrt{\rho}
\end{equation}

Here, $\mu$ is the shear modulus, $b$ is the magnitude of the Burgers vector and $\alpha$ is an empirical coefficient which takes values between 0.3 and 0.5. The dislocation density is composed of the sum of the density $\rho_S$ for statistically stored dislocations and the density $\rho_G$ for geometrically necessary dislocations as

\begin{equation}\label{Eq2MSG}
\rho = \rho_S + \rho_G
\end{equation}

The GND density $\rho_G$ is related to the effective plastic strain gradient $\eta^{p}$ by: 

\begin{equation}\label{Eq3MSG}
\rho_G = \overline{r}\frac{\eta^{p}}{b}
\end{equation}

\noindent where $\overline{r}$ is the Nye-factor which is assumed to be 1.90 for face-centered-cubic (fcc) polycrystals. Following \citet{FH97}, \citet{G99} used three quadratic invariants of the plastic strain gradient tensor to represent the effective plastic strain gradient $\eta^{p}$ as

\begin{equation}
\eta^{p}=\sqrt{c_1 \eta^{p}_{iik} \eta^{p}_{jjk} + c_2 \eta^{p}_{ijk} \eta^{p}_{ijk} + c_3 \eta^{p}_{ijk} \eta^{p}_{kji}}
\end{equation}

The coefficients were determined to be equal to $c_1=0$, $c_2=1/4$ and $c_3=0$ from three dislocation models for bending, torsion and void growth, leading to

\begin{equation}
\eta^{p}=\sqrt{\frac{1}{4}\eta^{p}_{ijk} \eta^{p}_{ijk}}
\end{equation}

\noindent where the components of the strain gradient tensor are obtained by $\eta^{p}_{ijk}= \varepsilon^{p}_{ik,j}+\varepsilon^{p}_{jk,i}-\varepsilon^{p}_{ij,k}$.

The tensile flow stress $\sigma_{flow}$ is related to the shear flow stress $\tau$ by:

\begin{equation}\label{Eq4MSG}
\sigma_{flow} =M\tau
\end{equation}

\noindent where $M$ is the Taylor factor taken to be 3.06 for fcc metals. Rearranging Eqs. (\ref{Eq1MSG}-\ref{Eq3MSG}) and Eq. (\ref{Eq4MSG}) yields

\begin{equation}\label{Eq5MSG}
\sigma_{flow} =M\alpha \mu b \sqrt{\rho_{S}+\overline{r}\frac{\eta^{p}}{b}}
\end{equation}

The SSD density $\rho_{S}$ can be determined from (\ref{Eq5MSG}) knowing the relation in uniaxial tension between the flow stress and the material stress-strain curve as follows

\begin{equation}\label{Eq6MSG}
\rho_{S} = [\sigma_{ref}f(\varepsilon^{p})/(M\alpha \mu b)]^2
\end{equation}

Here $\sigma_{ref}$ is a reference stress and $f$ is a non-dimensional function of the plastic strain $\varepsilon^{p}$ determined from the uniaxial stress-strain curve. Substituting back into (\ref{Eq5MSG}), $\sigma_{flow}$ yields:

\begin{equation}\label{EqSflow}
\sigma_{flow} =\sigma_{ref} \sqrt{f^2(\varepsilon^{p})+l\eta^{p}}
\end{equation}

\noindent where $l$ is the intrinsic material length based on parameters from of elasticity ($\mu$), plasticity ($\sigma_{ref}$) and atomic spacing ($b$):

\begin{equation}\label{Eqell}
l=M^2\overline{r}\alpha^2 \left(\frac{\mu}{\sigma_{ref}}\right)^2b=18\alpha^2\left(\frac{\mu}{\sigma_{ref}}\right)^2b
\end{equation}

\section{Finite element implementation}
\label{FE framework}

\subsection{Phenomenological approach}
\label{Fleck and Hutchinson's gradient theory}

A finite strain version of the gradient theory by \citet{FH01} is implemented following the work of \citet{NR04}, where a thorough description can be found (see also \citealp{NT05}). An updated Lagrangian configuration is adopted and by means of Kirchhoff stress measures the incremental principle of virtual work, Eq. (\ref{PVW}), can be expressed as:

\begin{align}\label{Eq:PVW2}
& \int_V \left( \accentset{\triangledown}{\varsigma}_{ij} \delta \dot{\varepsilon}_{ij} - \sigma_{ij} \left(2 \dot{\varepsilon}_{ik} \delta \dot{\varepsilon}_{kj} - \dot{e}_{kj} \delta \dot{e}_{ki} \right)+ \left(\dot{q} - \dot{\sigma}_e^{\varsigma} \right)\delta \dot{\varepsilon}^p + \accentset{\vee}{\varrho}_{i} \delta \dot{\varepsilon}_{,i}^p \right) \textnormal{dV} \nonumber \\
&= \int_S \left( \dot{T}_{0i} \delta \dot{u}_i + \dot{t}_0 \delta \dot{\varepsilon}^p \right)\textnormal{dS}
\end{align}

Here, $\accentset{\triangledown}{\varsigma}_{ij}$ is the Jaumann rate of the Kirchhoff stress, $\dot{q}$ is the rate of the Kirchhoff variant of the effective stress, $\accentset{\vee}{\varrho}_{i}$ is the convected derivative of the higher order Kirchhoff stress and the velocity gradient is denoted by $\dot{e}_{ij}$. $\dot{T}_{0i}$ and $\dot{t}_0$ are the nominal traction and the nominal higher order traction, respectively, with the subscript 0 referring to the reference configuration. The Kirchhoff quantities are related to their Cauchy counterparts in Eq. (\ref{PVW}) by the determinant, $J$, of the deformation gradient: $\varsigma_{ij}=J \sigma_{ij}$, $\varrho_i=J \zeta_i$, $q=JQ$ and $\sigma_e^{\varsigma}=J \sigma_e$. The finite strain generalization, for a hardening modulus $h\left[E^p\right]$, of the constitutive equations for the stress measures corresponding to the total strain, the plastic strain, and the plastic strain gradient, respectively, are given by:

\begin{equation}
\accentset{\triangledown}{\varsigma}_{ij} = \mathscr{D}_{ijkl} \left(\dot{\varepsilon}_{kl} - \dot{\varepsilon}^p m_{kl} \right)=\dot{\varsigma}_{ij} - \dot{\omega}_{ik} \sigma_{kj} - \sigma_{ik} \dot{\omega}_{jk}
\end{equation}

\begin{equation}
\dot{q}-\dot{\sigma}^{\varsigma}_{(e)}=h \left( \dot{\varepsilon}^p + \frac{1}{2} B_i \dot{\varepsilon}_{,i}^p +C \dot{\varepsilon}^p \right) - m_{ij} \accentset{\triangledown}{\varsigma}_{ij}
\end{equation}

\begin{equation}
\accentset{\vee}{\varrho}_{i}=h \left( A_{ij} \dot{\varepsilon}_{,j}^p + \frac{1}{2} B_i \dot{\varepsilon}^p \right)= \dot{\varrho}_i - \dot{e}_{ik} \varrho_k
\end{equation}

\noindent where the elastic stiffness tensor is given by

\begin{equation}
\mathscr{D}_{ijkl}=\frac{E}{1+\nu} \left( \frac{1}{2} \left( \delta_{ik} \delta_{jl} + \delta_{il} \delta_{jk} \right) + \frac{\nu}{1-2\nu} \delta_{ij} \delta_{kl} \right)
\end{equation}

\noindent and $\dot{\omega}_{ij}=\frac{1}{2} \left(\dot{e}_{ij}- \dot{e}_{ji} \right)$ is the anti-symmetric part of the velocity gradient. Here $\delta_{ij}$ is the Kronecker delta while $E$ and $\nu$ denote Young's modulus and the Poisson ratio, respectively.\\

A special kind of finite element (FE) method is used where, in addition to the nodal displacement increments, $\dot{U}^n$, the nodal effective plastic strain increments, $\dot{\varepsilon}^p_n$, appear directly as unknowns. The displacement increments, $\dot{u}_i$, and the effective plastic strain increments, $\dot{\varepsilon}^p$, are interpolated within each element by means of the shape functions:

\begin{equation}\label{Eq:FEinter}
\dot{u}_i= \sum_{n=1}^{2k_u} N_i^n \dot{U}^n \:, \; \; \; \; \; \; \; \dot{\varepsilon}^p=\sum_{n=1}^{k_p} M^n \dot{\varepsilon}^{p^n}
\end{equation}

\noindent where $k_u$ and $k_p$ are the number of nodes used for the displacement and effective plastic strain interpolations, respectively. The components $N_i^n$ $(i=1,2; \, n=1,...,2k_u)$ form the shape function matrix which by multiplication with the array $\dot{U}^n (n=1,...,2k)$ gives the displacement field. Similarly, the equivalent plastic strain field is determined from the shape function matrix $M^n$ and the array of nodal effective plastic strain increments $\dot{\varepsilon}^{p^n}$. By introducing the FE interpolation of the displacement field and the effective plastic strain field (\ref{Eq:FEinter}), and their appropriate derivatives, in the principle of virtual work (\ref{Eq:PVW2}), the following discretized system of equations is obtained:

\begin{equation}\label{Eq:Dis}
\begin{bmatrix}
  \boldsymbol{K_e} & \boldsymbol{K_{ep}} \\
  \boldsymbol{K_{ep}^T} & \boldsymbol{K_p}
 \end{bmatrix}
\begin{bmatrix}
\dot{\boldsymbol{U}} \\
\dot{\boldsymbol{\varepsilon}}^{\boldsymbol{p}} 
\end{bmatrix}=
\begin{bmatrix}
\dot{\boldsymbol{F}}_{\boldsymbol{1}}\\
\dot{\boldsymbol{F}}_{\boldsymbol{2}}
\end{bmatrix}
\end{equation}

Here, $\boldsymbol{K_e}$ is the elastic stiffness matrix, $\boldsymbol{K_{ep}}$ is a coupling matrix of dimension force and $\boldsymbol{K_p}$ is the plastic resistance, a matrix of dimension energy. The first part of the right-hand side of Eq. (\ref{Eq:Dis}) is composed of the conventional external incremental force vector $\dot{\boldsymbol{F}}_{\boldsymbol{1}}$ and the incremental higher order force vector $\dot{\boldsymbol{F}}_{\boldsymbol{2}}$. In the elastic regime the plastic strain contribution is disabled by setting $\boldsymbol{K_{ep}}=0$ and the weight of $\boldsymbol{K_p}$ is minimized by multiplying it by a small factor (e.g. $10^{-8}$), preserving the non-singular nature of the global system. The latter numerical feature eliminates any significant contribution to the solution of the nodal plastic strain increments on the current elastic-plastic boundary. This lack of constraint of plastic flow at the internal boundary can be physically interpreted as allowing dislocations to pass through it, as is the case in conventional plasticity (for details see \citealp{NH03}).\\ 

Based on a forward Euler scheme, when nodal displacement and effective plastic strain increments have been determined, the updated strains, $\varepsilon_{ij}$, stresses, $\sigma_{ij}$, higher order stresses, $\zeta_i$, and $Q$ are computed at each integration point. Initial plastic yielding is initiated when $\sigma_e$ becomes larger than the initial yield stress $\sigma_y$. A time increment sensitivity analysis has been conducted in all computations to ensure that the numerical solution does not drift away from the correct one.\\

\subsection{Mechanism-based approach}
\label{Gao et al. theory}

\citet{H04} used a viscoplastic formulation to construct the conventional theory of mechanism-based strain gradient (CMSG) plasticity from the \citet{T38} dislocation model (see details in \citealp{H04}). In CMSG plasticity the plastic strain gradient comes into play through the incremental plastic modulus and therefore it does not involve higher order terms. The CMSG theory is chosen as it does not suffer convergence problems in large strains crack tip analysis, unlike its higher order counterpart: The finite deformation theory of MSG plasticity (see \citealp{H03} and \citealp{MB15}). The viscoplastic-limit approach developed by \citet{KO02} is employed to suppress strain rate and time dependence by replacing the reference strain rate $\dot{\varepsilon}_0$ with the effective strain rate $\dot{\varepsilon}$ in the viscoplastic-like power law adopted:

\begin{equation}
\dot{\varepsilon}^{p} = \dot{\varepsilon} \left [\frac{\sigma_e}{\sigma_{ref} \sqrt{f^{2}(\varepsilon^{p})+l\eta^{p}}} \right]^{m}
\end{equation}

The exponent is taken to fairly large values ($m\geq20$) which in\citet{KO02} scheme is sufficient to reproduce the rate-independent behavior given by the viscoplastic limit in a conventional power law (see \citealp{H04}). Taking into account that the volumetric and deviatoric strain rates are related to the stress rate in the same way as in classical plasticity, the constitutive equation yields:

\begin{equation}
\dot{\sigma}_{ij}=K\dot{\varepsilon}_{kk}\delta_{ij}+2\mu \left\{\dot{\varepsilon}'_{ij} - \frac{3\dot{\varepsilon}}{2\sigma_e}\left[\frac{\sigma_e}{\sigma_{flow}} \right]^{m} \dot{s}_{ij} \right\}
\end{equation}

Here $K$ being the bulk modulus. As it is based on the Taylor dislocation model, which represents an average of dislocation activities, the CMSG theory is only applicable at a scale much larger than the average dislocation spacing. For common values of dislocation density in metals, the lower limit of physical validity of MSG plasticity is around 100 nm. Although higher order terms are required to model effects of dislocation blockage at impermeable boundaries (see \citealp{NH03b}), one should note that higher order boundary conditions have essentially no effect on the stress distribution at a distance of more than 10 nm away from the crack tip in MSG plasticity \citep{S01,Q04}, well below its lower limit of physical validity.\\

Since higher order terms are not involved, the governing equations of CMSG plasticity are essentially the same as those in conventional plasticity and the FE implementation is quite straightforward. The plastic strain gradient is obtained by numerical differentiation within the element: the plastic strain increment is interpolated through its values at the Gauss points in the isoparametric space and afterwards the increment in the plastic strain gradient is calculated by differentiation of the shape functions. Rigid body rotations for the strains and stresses are carried out by means of the \citet{HW80} algorithm and the strain gradient is obtained from the deformed configuration since the infinitesimal displacement assumption is no longer valid (see \citealp{MB15}).

\section{Numerical results}
\label{FE Results}

\subsection{Infinitesimal deformation theory}

Results obtained for small strains will allow us to introduce the comparative study between theories and to validate the present numerical implementation with results obtained from the literature. Two dimensional plane strain crack tip fields are evaluated by means of a boundary layer formulation, where the crack region is contained by a circular zone and the Mode I load is applied at the remote circular boundary through a prescribed displacement: 

\begin{equation}
u(r,\theta)=K_I \frac{1+\nu}{E} \sqrt{\frac{r}{2\pi}}cos\left(\frac{\theta}{2}\right)(3-4\nu-cos\theta)
\end{equation}

\begin{equation}
v(r,\theta)=K_I \frac{1+\nu}{E} \sqrt{\frac{r}{2\pi}}sin\left(\frac{\theta}{2}\right)(3-4\nu-cos\theta)
\end{equation}

\begin{figure}[H]
        \centering
        \begin{subfigure}[h]{0.45\textwidth}
                \centering
                \includegraphics[scale=0.75]{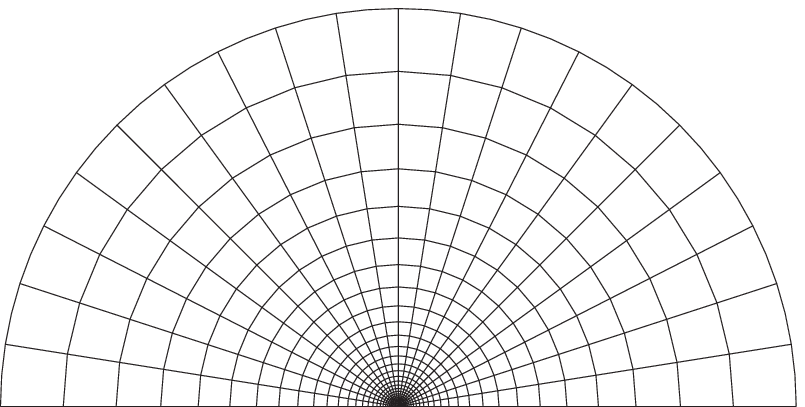}
                \caption{}
                \label{fig:Fig1a}
        \end{subfigure}
        \begin{subfigure}[h]{0.49\textwidth}
                \centering
                \includegraphics[scale=0.4]{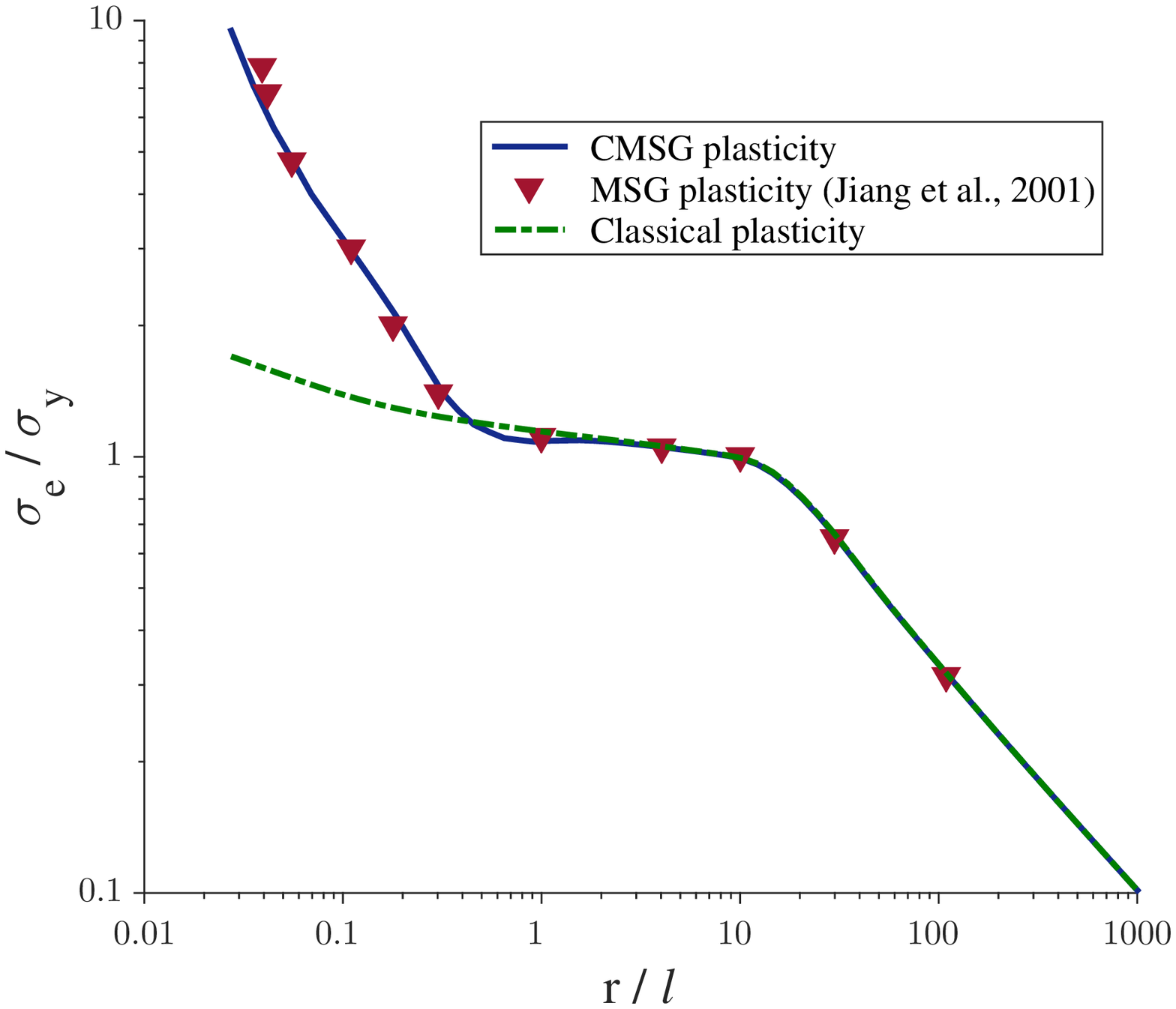}
                \caption{}
                \label{fig:Fig1b}
        \end{subfigure}
       
        \caption{(a) Finite element mesh for the boundary layer formulation; (b) Comparison between MSG and CMSG predictions}\label{fig:Fig1}
\end{figure}

Here, $u$ and $v$ are the horizontal and vertical components of the displacement boundary condition, $r$ and $\theta$ the radial and angular coordinates in a polar coordinate system centered at the crack tip, $E$ is Young's modulus and $\nu$ is the Poisson ratio of the material and $K_I$ is the applied stress intensity factor, which quantifies the remote load. Plane strain conditions are assumed and only the upper half of the circular domain is modeled due to symmetry. An outer radius of $R=42 \, mm$ is defined and the entire specimen is discretized by means of eight-noded quadrilateral elements with reduced integration. Different mesh densities were used to study convergence behavior, and it was found that 1600 elements were sufficient to achieve mesh-independent results. With the aim of accurately characterizing the influence of the strain gradient a very refined mesh is used near the crack tip, where the size of the elements is on the order of nanometers (see fig. \ref{fig:Fig1a}). Unless otherwise stated, the following set of non-dimensional material parameters is considered in the present work

\begin{equation}\label{Eq:prop}
N=0.2, \, \, \, \, \, \, \, \, \, \, \, \, \frac{\sigma_Y}{E}=0.2\%, \, \, \, \, \, \, \, \, \, \, \, \, \nu=0.3
\end{equation}

\noindent where $\sigma_Y$ is the initial yield stress and $N$ is the strain hardening exponent. An isotropic power law material is adopted according to

\begin{equation}
\sigma=\sigma_Y \left( 1 + \frac{E \varepsilon^p}{\sigma_Y} \right)^N
\end{equation}

In the phenomenological approach, the hardening curve is evaluated at $E^p$ instead of $\varepsilon^p$ as discussed in \citet{FH01}. The reference stress of (\ref{Eq6MSG}) will correspond to $\sigma_{ref}=\sigma_Y \left( \frac{E}{\sigma_Y}\right)^{N}$ and $f(\varepsilon^{p})=\left(\varepsilon^{p}+\frac{\sigma_Y}{E} \right)^{N}$. Fig. \ref{fig:Fig1b} shows, in a double logarithm diagram, the normalized effective stress $\sigma_e / \sigma_Y$ versus the normalized distance $r/l$ ahead of the crack tip ($\theta=1.014^{\circ}$) for an external applied load of $K_I=20 \sigma_Y \sqrt{l}$. As it can be seen in the figure, a very good agreement is obtained between the stress distributions obtained by means of the CMSG theory and MSG plasticity (taken from \citealp{J01}), showing that higher order boundary conditions do not influence crack tip fields within its physical domain of validity. Consequently, all the results obtained from the CMSG theory are henceforth labeled as MSG plasticity. Results prove the suitability of CMSG plasticity in the present study, allowing to develop a robust implicit numerical scheme (see \citealp{MB15}) \\

\newpage

\newpage

\begin{figure}[H]
\centering
\includegraphics[scale=0.85]{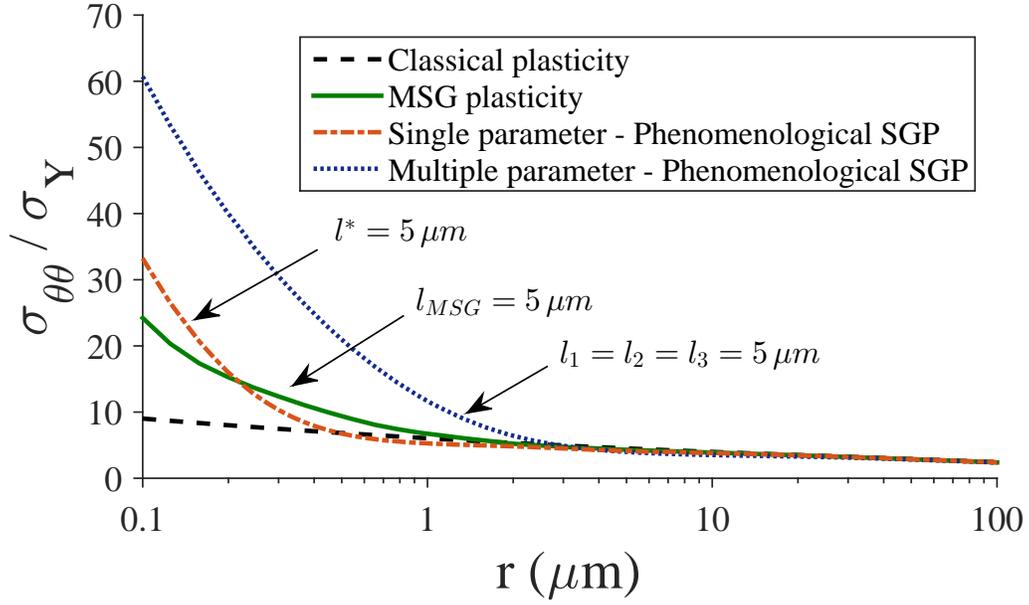}
\caption{Small strain predictions of $\sigma_{\theta\theta}$ ahead of the crack tip for classical plasticity and both mechanism-based and phenomenological SGP approaches. The figure shows results along the extended crack plane with the distance to the crack tip $r$ in log scale for $K_{I}=25\sigma_Y \sqrt{l}$, $\sigma_Y=0.2\%E$, $\nu=0.3$, $N=0.2$ and material length scales of $l^{*}=l_1=l_2=l_3=l_{MSG}=5$ $\mu m$}
\label{fig:S22smallstrains}
\end{figure}

Fig. \ref{fig:S22smallstrains} shows the opening stress distributions $\sigma_{\theta\theta}$ ahead of the crack tip $(\theta=0^{\circ})$ obtained from classical plasticity, phenomenological SGP (both single length and multiple length parameter theories) and MSG plasticity. The stress values are normalized by the material yield stress while the horizontal axis is left unchanged, due to the central role that the magnitude of the domain ahead of the crack tip influenced by strain gradients plays on damage modeling. In the present study, a material length scale of $l=5$ $\mu m$ has been considered. This would be a typical estimate for nickel \citep{SE98} and corresponds to an intermediate value within the range of experimentally observed material length scales reported in the literature (1-10 $\mu m$).\\

\newpage

\newpage

Results show that SGP stress predictions agree with classical plasticity away from the crack tip but become much larger within tens of microns from it. Fig. \ref{fig:S22smallstrains} reveals significant quantitative differences among theories for the same reference value of the material length scale. Within the phenomenological approach, the single length scale theory predicts much smaller size effects than the multiple parameter theory when all individual length scales $l_i$ are set equal to $l^*$, as previously reported by \citet{K08}. Furthermore, it is seen that the stress level attained near the crack tip from the phenomenological approach is much higher than MSG plasticity predictions, especially in the case of the multiple length scale theory. However, the distance ahead of the crack tip where the stress distribution deviates from classical plasticity predictions is quite similar for the cases of MSG plasticity and the single parameter phenomenological theory, while a significantly larger size of the domain influenced by strain gradients is observed when the multiple length parameter theory is adopted. 

\subsection{Finite deformation theory}
\label{Results obtained for large strains}

\begin{figure}[H]
        \centering
        \begin{subfigure}[h]{0.45\textwidth}
                \centering
                \includegraphics{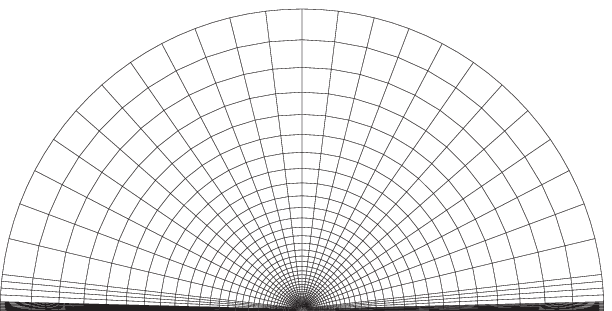}
                \caption{}
                \label{fig:FigMBLa}
        \end{subfigure}
        \begin{subfigure}[h]{0.45\textwidth}
                \centering
                \includegraphics{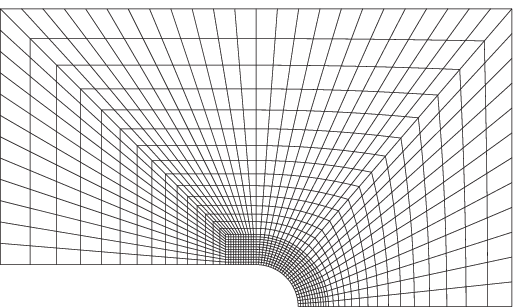}
                \caption{}
                \label{fig:FigMBLb}
        \end{subfigure}
       
        \caption{Finite element mesh for the boundary layer formulation under large deformations: (a) complete model and (b) vicinity of the crack}\label{fig:FigMBL}
\end{figure}

Since large strains take place in the vicinity of the crack, crack tip fields should be evaluated within the framework of the finite deformation theory in order to assess the influence of strain gradients in damage and fracture modeling. Moreover, the results of \citet{MB15} reveal a meaningful increase in the domain influenced by the size effect when large strains are taken into account, as a consequence of the influence of strain gradients on the work hardening of the material. The initial configuration and the background mesh of the boundary layer formulation are shown in fig. \ref{fig:FigMBL}. Following \citet{M77}, a ratio between the radii of the outer boundary and the crack tip of $R/r=10^5$ is considered and, as in the small strain case, different mesh densities were evaluated in order to compute accurate results. Around 6200 eight-noded quadrilateral elements with reduced integration were generally used to achieve convergence.

\newpage
\newpage

\begin{figure}[H]
\centering
\includegraphics[scale=0.85]{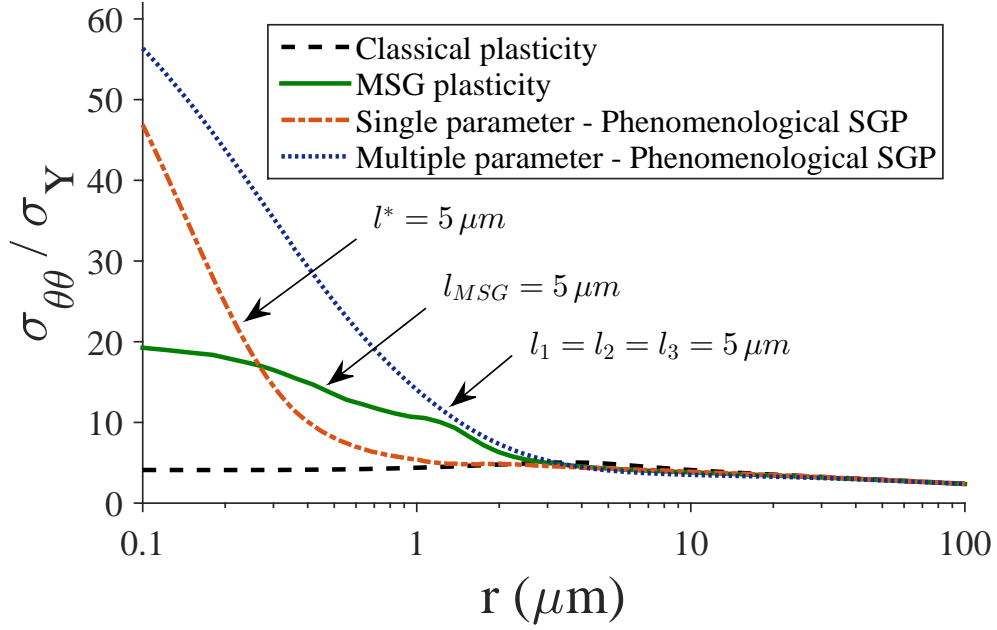}
\caption{Finite deformation results for $\sigma_{\theta\theta}$ ahead of the crack tip for classical plasticity and both mechanism-based and phenomenological SGP approaches. The figure shows results along the extended crack plane with the distance to the crack tip $r$ in log scale for $K_{I}=25\sigma_Y \sqrt{l}$, $\sigma_Y=0.2\%E$, $\nu=0.3$, $N=0.2$ and material length scales of $l^{*}=l_1=l_2=l_3=l_{MSG}=5$ $\mu m$.}
\label{fig:S22largestrains}
\end{figure}

Fig. \ref{fig:S22largestrains} plots the normalized opening stress distribution under the same conditions as fig. \ref{fig:S22smallstrains} where, as in the small strains case, the distance to the crack tip $r$ is shown in logarithmic scale. Results obtained with classical plasticity reproduce the well known behavior revealed by \citet{M77}, namely that large strains at the crack tip cause the crack to blunt, reducing the stress triaxiality locally. However, when size effects are included in the modelization, strain gradients increase the resistance to plastic deformation, lowering crack tip blunting and consequently, suppressing the local stress reduction. As it can be seen in the figure, a monotonic stress increase is still observed in SGP predictions and therefore the distance ahead of the crack tip where the strain gradients severely influence the stress distributions increases significantly when compared to the small strain results.\\

\newpage
\newpage

As in the small strain case, results shown in fig. \ref{fig:S22largestrains} also reveal significant quantitative differences among SGP theories for the same reference material length scale. As in fig. \ref{fig:S22smallstrains}, the single length parameter phenomenological theory predicts a smaller influence of GNDs when compared to the multiple parameter version, although the magnitude of stress elevation computed close to the crack tip from both theories is much closer when finite strains are taken into account. Both single and multiple length scale phenomenological theories predict much higher stress levels at the crack tip than MSG plasticity. However, the domain ahead of the crack tip where size effects alter the stress distribution in MSG plasticity is significantly greater in finite strains, close to the predictions obtained from the Fleck-Hutchinson multiple length parameter theory for the load level considered. \\

\newpage
\begin{figure}[H]
\centering
\includegraphics[scale=0.85]{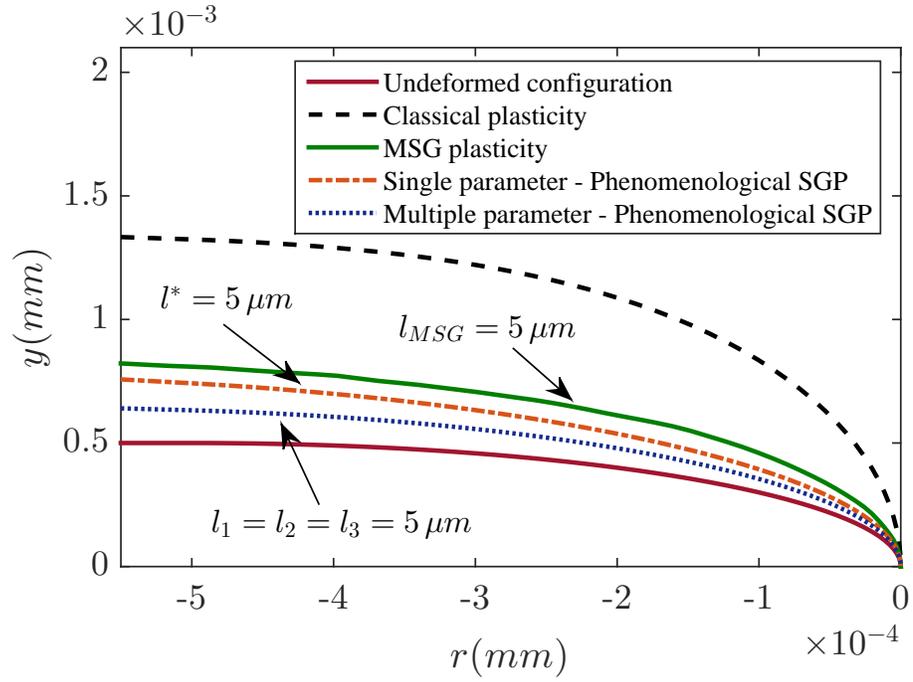}
\caption{Initial and final crack tip blunting predicted by classical plasticity and both mechanism-based and phenomenological SGP approaches for $K_{I}=25\sigma_Y \sqrt{l}$, $\sigma_Y=0.2\%E$, $\nu=0.3$, $N=0.2$ and material length scales of $l^{*}=l_1=l_2=l_3=l_{MSG}=5$ $\mu m$.}
\label{fig:CTOD}
\end{figure}
\newpage

Unlike classical plasticity, for all SGP stress distributions the maximum level of stress is achieved at the crack tip as a consequence of local hardening promoted by GNDs. Fig. \ref{fig:CTOD} shows the degree of crack tip blunting under the same conditions as fig. \ref{fig:S22largestrains} where it can be seen that blunting of the initial crack tip radius decreases significantly when size effects are included in the modelization. As the influence of strain gradients on crack tip fields persists all the way to the crack tip, essential differences arise when comparing with classical plasticity predictions in the blunting dominated zone. Hence, the magnitude of macroscopic stress elevation is much higher than that reported by previous studies, conducted within infinitesimal deformation theory.\\

\begin{figure}[h]
\makebox[\linewidth][c]{%
        \begin{subfigure}[b]{0.55\textwidth}
                \centering
                \includegraphics[scale=0.49]{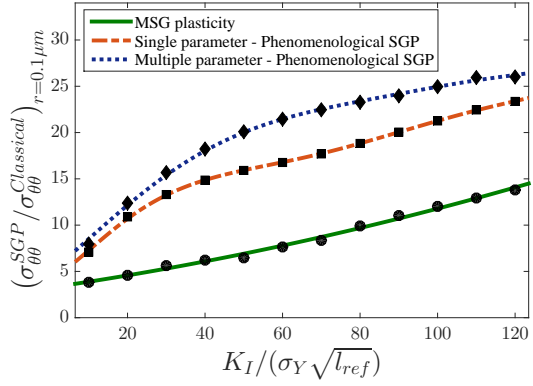}
                \caption{}
                \label{fig:Fig6a}
        \end{subfigure}
        \begin{subfigure}[b]{0.55\textwidth}
                \centering
                \includegraphics[scale=0.49]{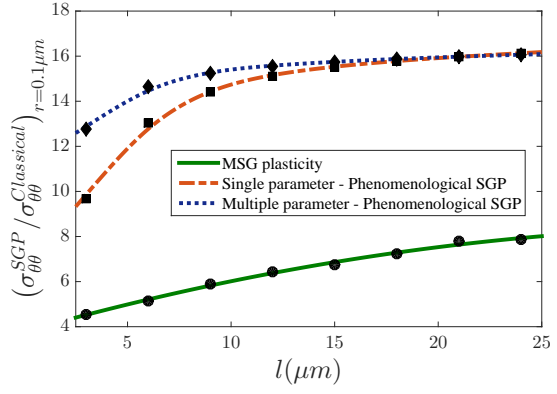}
                \caption{}
                \label{fig:Fig6b}
        \end{subfigure}
        }
       
        \caption{Ratio of stress elevation promoted by strain gradients at $r=0.1 \, \mu m$ ahead of the crack tip ($\theta=0^{\circ}$) as a function of (a) applied load $K_I$ and (b) material length scale $l$, for $\sigma_Y=0.2\%E$, $\nu=0.3$ and $N=0.2$. The length parameters in (a) are $l^{*}=l_1=l_2=l_3=l_{MSG}=5$ $\mu m$ while the reference applied load in (b) is $K_{I}=25\sigma_Y \sqrt{l_{ref}}$ (with $l_{ref}=5 \, \mu m$)}\label{fig:Fig6}
\end{figure}

\begin{figure}[H]
\makebox[\linewidth][c]{%
        \begin{subfigure}[b]{0.55\textwidth}
                \centering
                \includegraphics[scale=0.49]{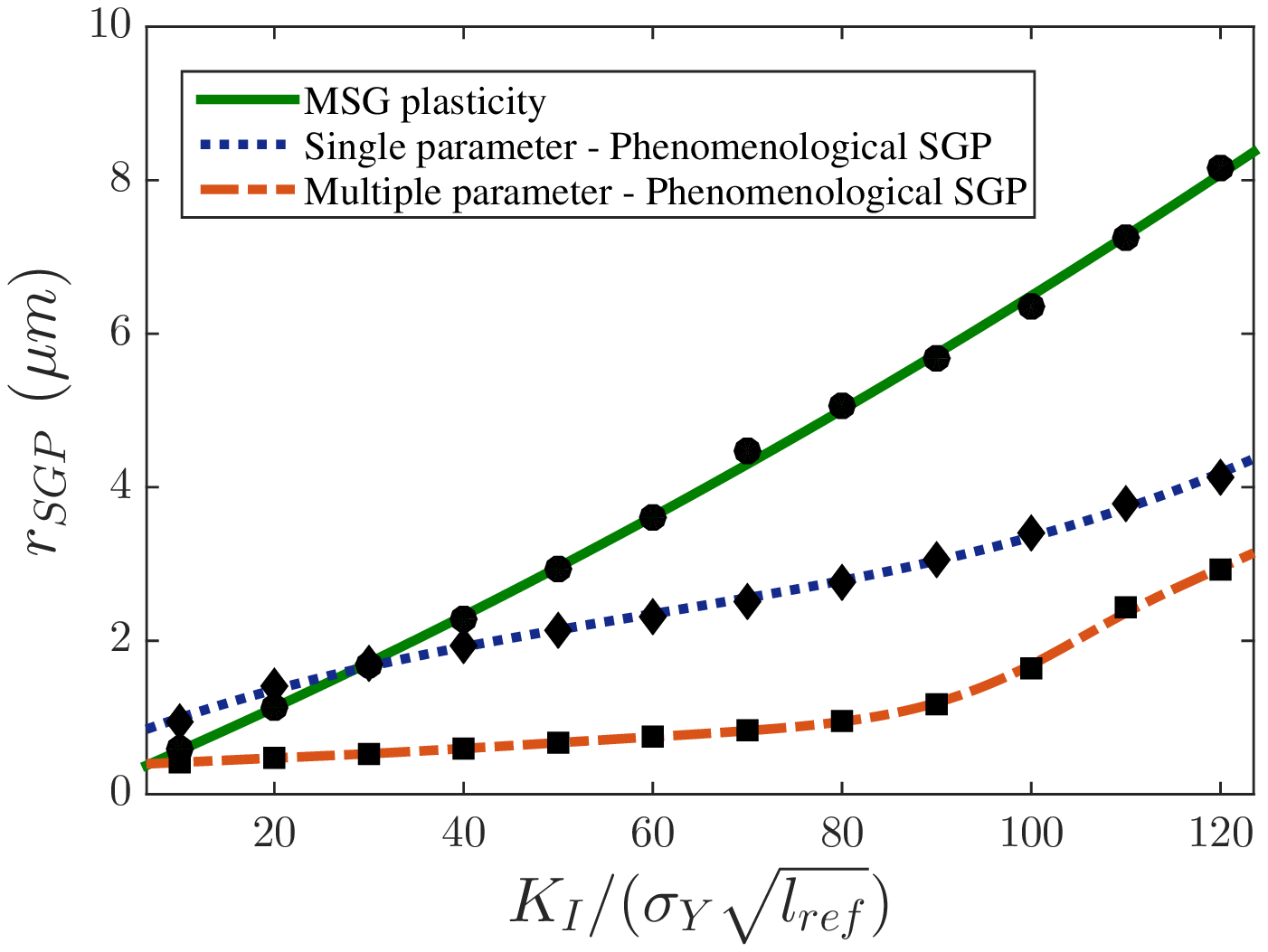}
                \caption{}
                \label{fig:Fig7a}
        \end{subfigure}
        \begin{subfigure}[b]{0.55\textwidth}
                \centering
                \includegraphics[scale=0.49]{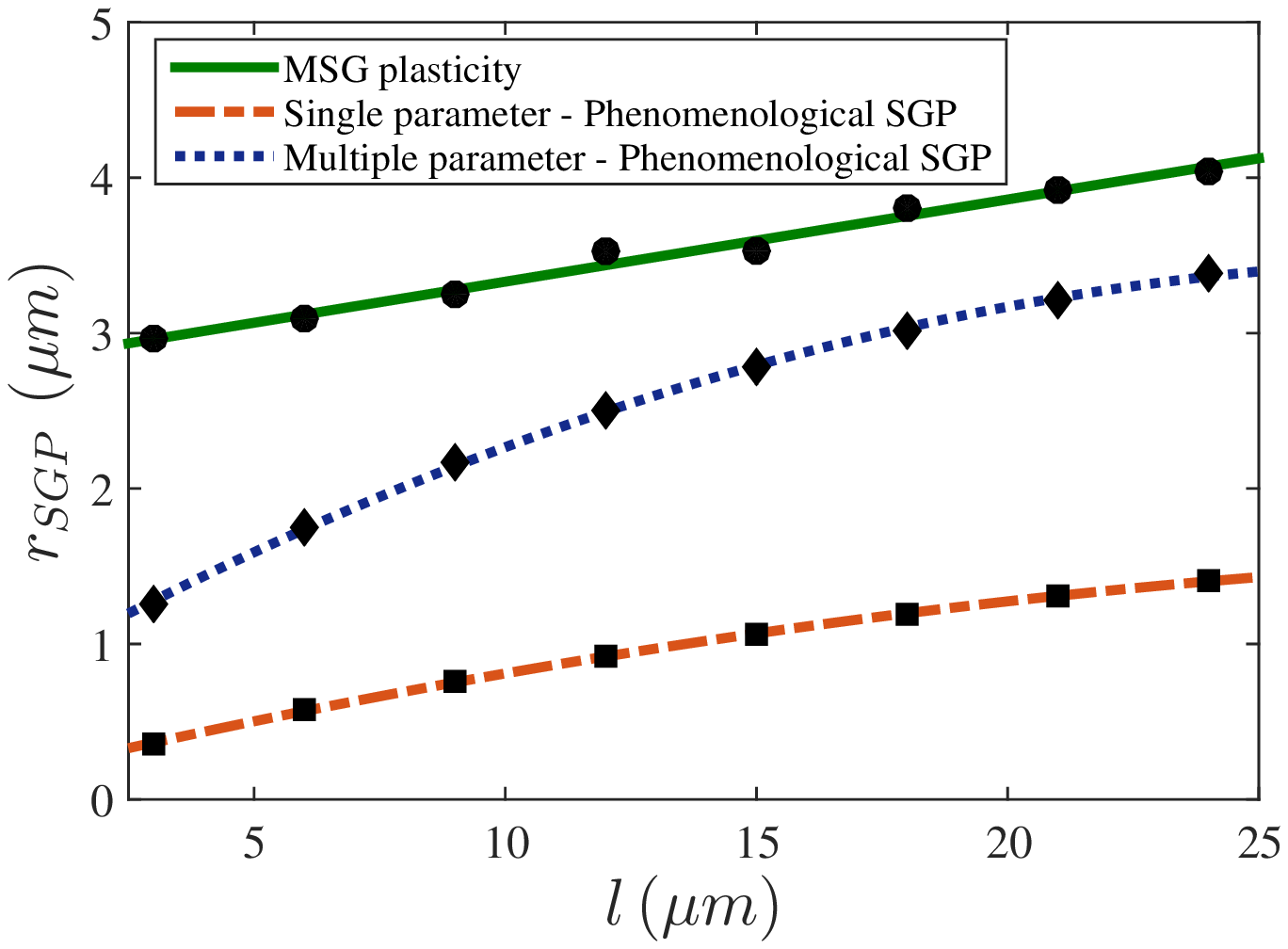}
                \caption{}
                \label{fig:Fig7b}
        \end{subfigure}
        }
       
        \caption{Distance ahead of the crack tip where the strain gradients significantly influence the stress distribution $r_{SGP}$ as a function of (a) applied load $K_I$ and (b) material length scale $l$, for $\sigma_Y=0.2\%E$, $\nu=0.3$ and $N=0.2$. The length parameters in (a) are $l^{*}=l_1=l_2=l_3=l_{MSG}=5$ $\mu m$ while the reference applied load in (b) is $K_{I}=25\sigma_Y \sqrt{l_{ref}}$ (with $l_{ref}=5 \, \mu m$)}\label{fig:Fig7}
\end{figure}

Figs. \ref{fig:Fig6} and \ref{fig:Fig7} quantify the differences from classical plasticity predictions as a function of (a) the external load and (b) the material length scale. Both the magnitude of stress elevation close to the crack tip and the physical length over which gradient effects significantly enhance crack tip stresses are evaluated. The figs. \ref{fig:Fig6} and \ref{fig:Fig7} show, respectively, the variation of the ratio of stress elevation $\sigma_{SGP} / \sigma_{Classical}$ at $r=0.1 \, \mu m$ and $r_{SGP}$, the size of the domain ahead of the crack tip where the stress distribution significantly deviates from classical plasticity predictions $\left(\sigma_{SGP} > 2 \sigma_{Classical}\right)$. In Fig. 6 stresses are sampled at $r=0.1 \, \mu m$ as it is considered the lower limit of physical validity of continuum SGP theories, while being sufficiently close to the crack tip to provide representative results of interest for the modelization of several damage mechanisms. \\

\newpage
\newpage

\newpage

\begin{figure}[H]
\centering
\includegraphics[scale=0.95]{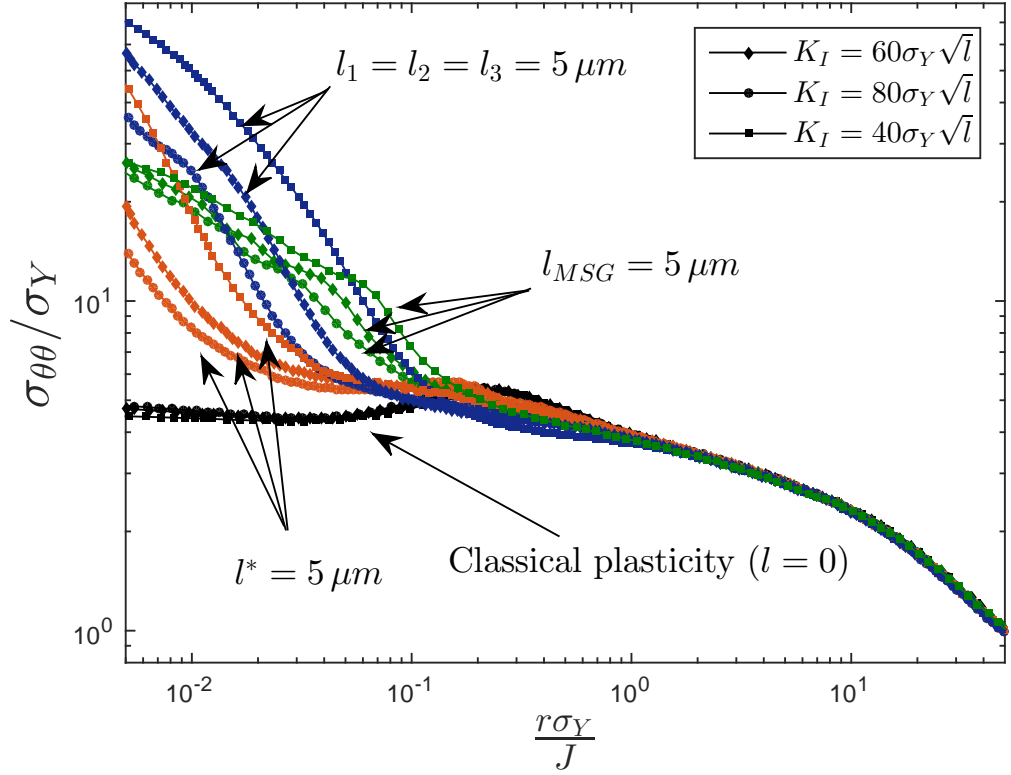}
\caption{Double logarithm plot of the normalized opening stress distribution $\sigma_{\theta\theta} / \sigma_Y$ ahead of the crack tip for classical plasticity and both mechanism-based and phenomenological SGP approaches, being the distance to the crack tip normalized by the external load $r \sigma_Y / J$ for $\sigma_Y=0.2\%E$, $\nu=0.3$, $N=0.2$ and material length scales of $l^{*}=l_1=l_2=l_3=l_{MSG}=5$ $\mu m$. Finite deformation theory}
\label{fig:LoadNorm}
\end{figure}

In both phenomenological and mechanism-based approaches the magnitude of stress elevation and the domain of influence of strain gradients monotonically increase with the external load and the value of the reference length scale parameter. For the higher load level considered the opening stress value at the crack tip is 15-25 times the estimation of classical plasticity, depending on the SGP theory considered, while the distance ahead of the crack where strain gradients significantly alter stress distributions spans several micrometers. One should note that a wide range of load levels of interest for damage modeling has been considered, with the largest load level roughly $K_I \approx 100$ MPa$\sqrt{m}$ for a typical steel of $\sigma_Y=400$ MPa and $E=200000$ MPa. Both the domain influenced by strain gradients and the ratio of stress elevation at the crack tip show sensitivity to the length scale parameter, especially for lower values of $l$. In fact, for high values of $l$ both MSG plasticity and the phenomenological multiple length parameter theory predict an SGP influenced region bigger than the blunting dominated zone. Thus, for some particular combinations of $l$, applied load and material properties, the physical length over which strain gradients meaningfully enhance crack tip stresses spans several tens of micrometers. This may have important implications on fracture and damage modeling of metals since the critical distance of many damage mechanisms fall within this range. Moreover, damage modelization at the continuum level has been generally based on a distinct feature of classical plasticity: the peak stress ahead of the crack tip changes its position with the load but does not change its value. This is not the case when accounting for strain gradient effects in the constitutive modeling, as shown in fig. \ref{fig:LoadNorm}, where the normalized opening stress distribution $\sigma_{\theta \theta} / \sigma_Y$ ahead of the crack tip is shown in a double logarithmic plot for different values of the crack tip load.\\

\newpage
\newpage

The distance to the crack tip has been normalized by the external load $r \sigma_Y / J$, with $J$ denoting the $J$-integral, that is related to the applied load by $J=\left(1-\nu^2\right) K_I^2 / E$. The figure reveals that the influence of GNDs persists all the way to the crack tip, even for very large amounts of crack tip blunting. Unlike classical plasticity (represented by the black curves), crack tip fields obtained from SGP theories cannot be scaled by the load and the maximum stress level increases with the external load. \\

The present results highlight the need to account for the influence of strain gradients in the modelization of several damage mechanisms. The extent ahead of the crack tip where strain gradients play an important role suggests that gradient enhanced simulations may be relevant for continuum modeling of cleavage fracture \citep{Q11}, ductile-to-brittle assessment \citep{B08}, fatigue crack closure \citep{F86} and ductile damage \citep{G75,CN80,L05}. Furthermore, accounting for the influence of GNDs in the vicinity of the crack may be particularly relevant in the modelization of hydrogen assisted cracking, due to the essential role that the hydrostatic stress has on both interface decohesion and hydrogen diffusion in relation to the fracture process zone (see \citealp{G03}).  

\newpage
\begin{figure}[H]
\centering
\includegraphics[scale=0.85]{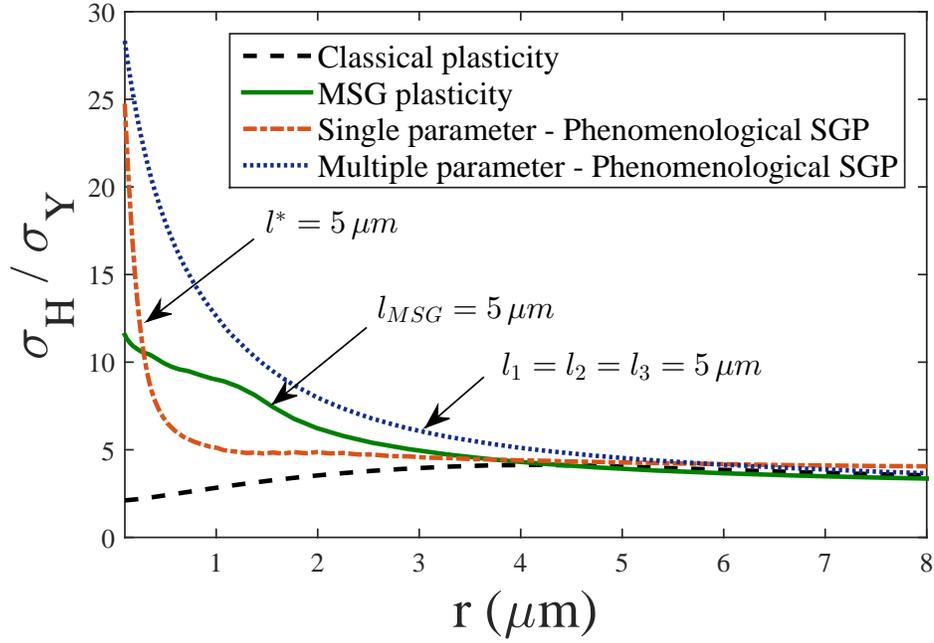}
\caption{Finite deformation theory results for $\sigma_H$ ahead of the crack tip for classical plasticity and both mechanism-based and phenomenological SGP approaches. The distance to the crack tip is denoted $r$ and the parameters of the problem are $K_{I}=25\sigma_Y \sqrt{l}$, $\sigma_Y=0.2\%E$, $\nu=0.3$, $N=0.2$ and and material length scales of $l^{*}=l_1=l_2=l_3=l_{MSG}=5$ $\mu m$.}
\label{fig:SHlargestrains}
\end{figure}
\newpage

Fig. \ref{fig:SHlargestrains} shows the hydrostatic stress distribution ahead of the crack tip under the same conditions as fig. \ref{fig:S22largestrains}. Results reveal that $\sigma_H$ shows broadly identical trends as the opening stress. The conventional plasticity solution agrees with SGP predictions far from the crack tip but significant differences arise within several micrometers of the crack tip as the stress level decreases in the blunting dominated zone for conventional plasticity. The high level of crack tip surface hydrogen measured in high-strength steels suggests that damage takes place within 1 $\mu$m of the crack surface (see \citealp{C00,G03}). The stress level attained at $r=1$ $\mu$m from MSG plasticity and single and multiple length parameter phenomenological theories is, respectively, $\approx$ 3.5, 2 and 5 times the prediction of classical plasticity. Since results have been obtained for a load level ($\approx 20$ MPa$\sqrt{m}$ for a typical steel) that could be considered a lower bound for damage modeling (see e.g. \citealp{G14}), accounting for the influence of GNDs close to the crack tip appears to be imperative in hydrogen embrittlement models.\\

However, the quantitative differences observed among SGP theories hinder gradient enhanced modeling. Both opening (\cref{fig:S22largestrains,fig:CTOD,fig:LoadNorm}) and hydrostatic stress distributions (fig. \ref{fig:SHlargestrains}) reveal substantial dissimilarities under the same reference length parameter. A qualitative agreement is found when examining the influence of the external load and the material length scale parameter for both phenomenological and mechanism-based SGP theories (\cref{fig:Fig6,fig:Fig7}), although relevant quantitative differences are appreciated. A much higher value of $l$ is needed in MSG plasticity to reach the crack tip stress predicted by means of both versions of Fleck-Hutchinson theory (fig. \ref{fig:Fig6b}) while the opposite is true when examining the distance ahead of the crack tip where the stress distribution deviates from classical plasticity predictions (fig. \ref{fig:Fig7b}). Under the same conditions as fig. \ref{fig:CTOD} a close degree of crack tip blunting is obtained by means of the following relation:

\begin{equation}
l_1=l_2=l_3 \approx \frac{1}{5} l_{MSG} \approx \frac{1}{2.5} l^*
\end{equation}

Using a cohesive zone model, \citet{WQ04} established that the relation between the steady-state fracture toughness and the separation strength obtained from MSG plasticity and from an earlier version of the Fleck-Hutchinson theory \citep{FH97,WH97} agrees if one considers the following approximate relation for the length scale parameter:

\begin{equation}
l_{MSG} \approx (4-5) l_{SG}
\end{equation}

Here, $l_{MSG}$ and $l_{SG}$ are the material length scales of the MSG theory and the \citet{FH97} phenomenological theory, respectively. This correlation is similar to the one elucidated by means of crack tip blunting in the present work. However, since the material length scale has to be determined from micro-tests, it is still uncertain if the experimentally obtained value of $l$ for MSG plasticity will be $4-5$ times its counterpart in Fleck-Hutchinson theory. In fact, similar values of $l$ have been obtained for polycrystalline copper from both approaches \citep{F94,NG98} and therefore further research is needed to provide an accurate quantitative assessment of the influence of GNDs at the crack tip.\\

\begin{figure}[H]
\makebox[\linewidth][c]{%
        \begin{subfigure}[b]{0.38\textwidth}
                \centering
                \includegraphics[scale=0.36]{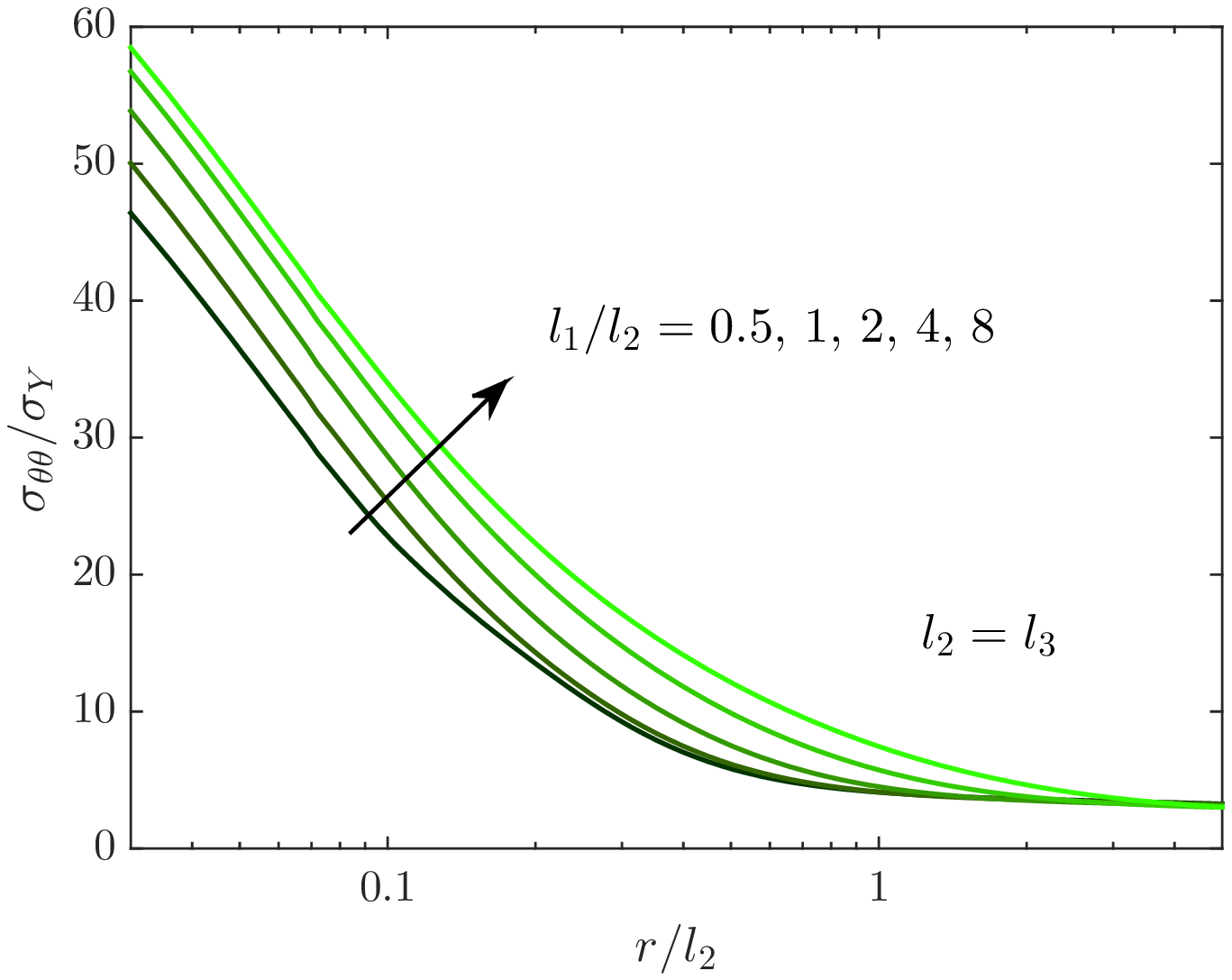}
                \caption{}
                \label{fig:Fig10a}
        \end{subfigure}
        \begin{subfigure}[b]{0.38\textwidth}
                \centering
                \includegraphics[scale=0.36]{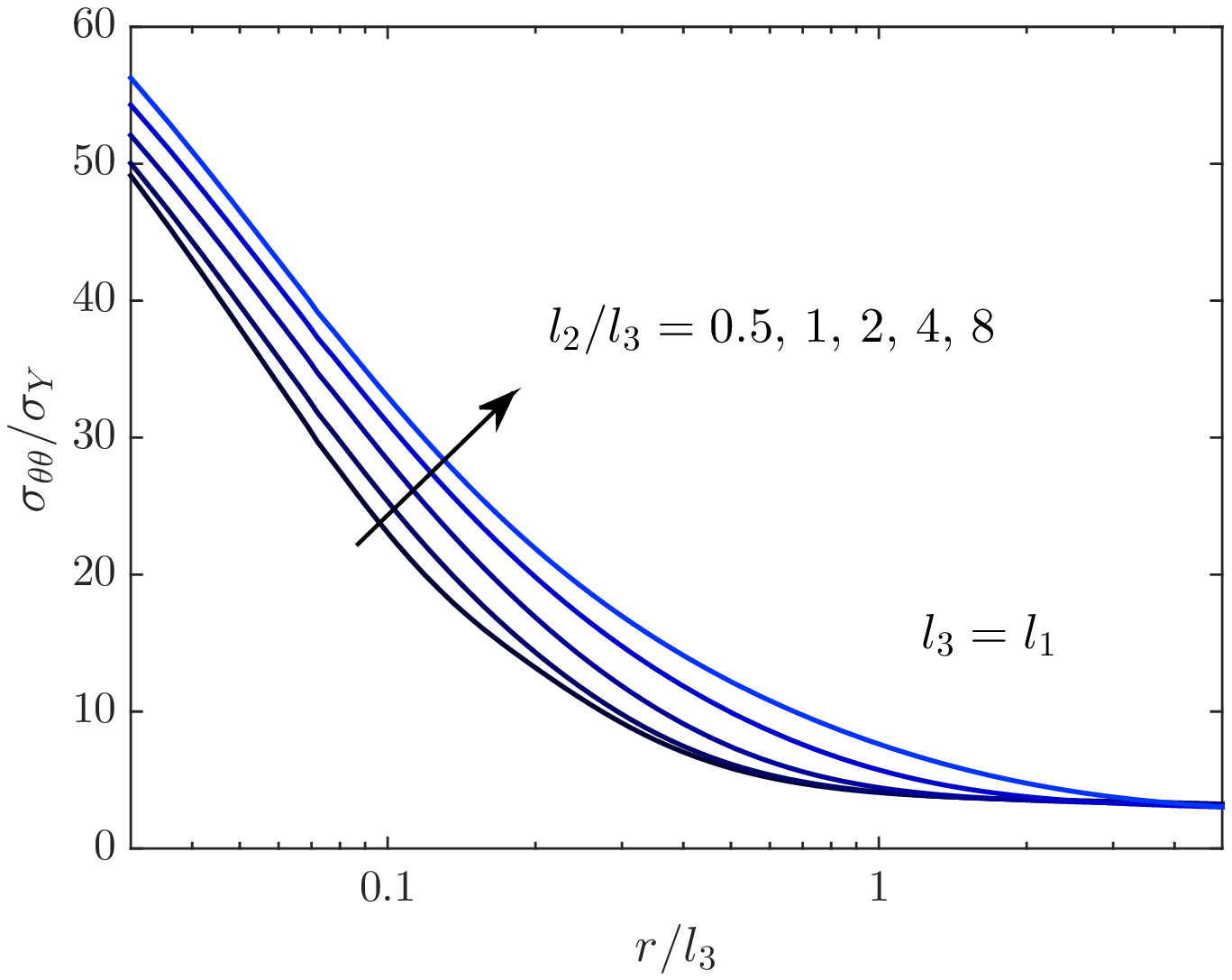}
                \caption{}
                \label{fig:Fig10b}
        \end{subfigure}
        \begin{subfigure}[b]{0.38\textwidth}
                \centering
                \includegraphics[scale=0.36]{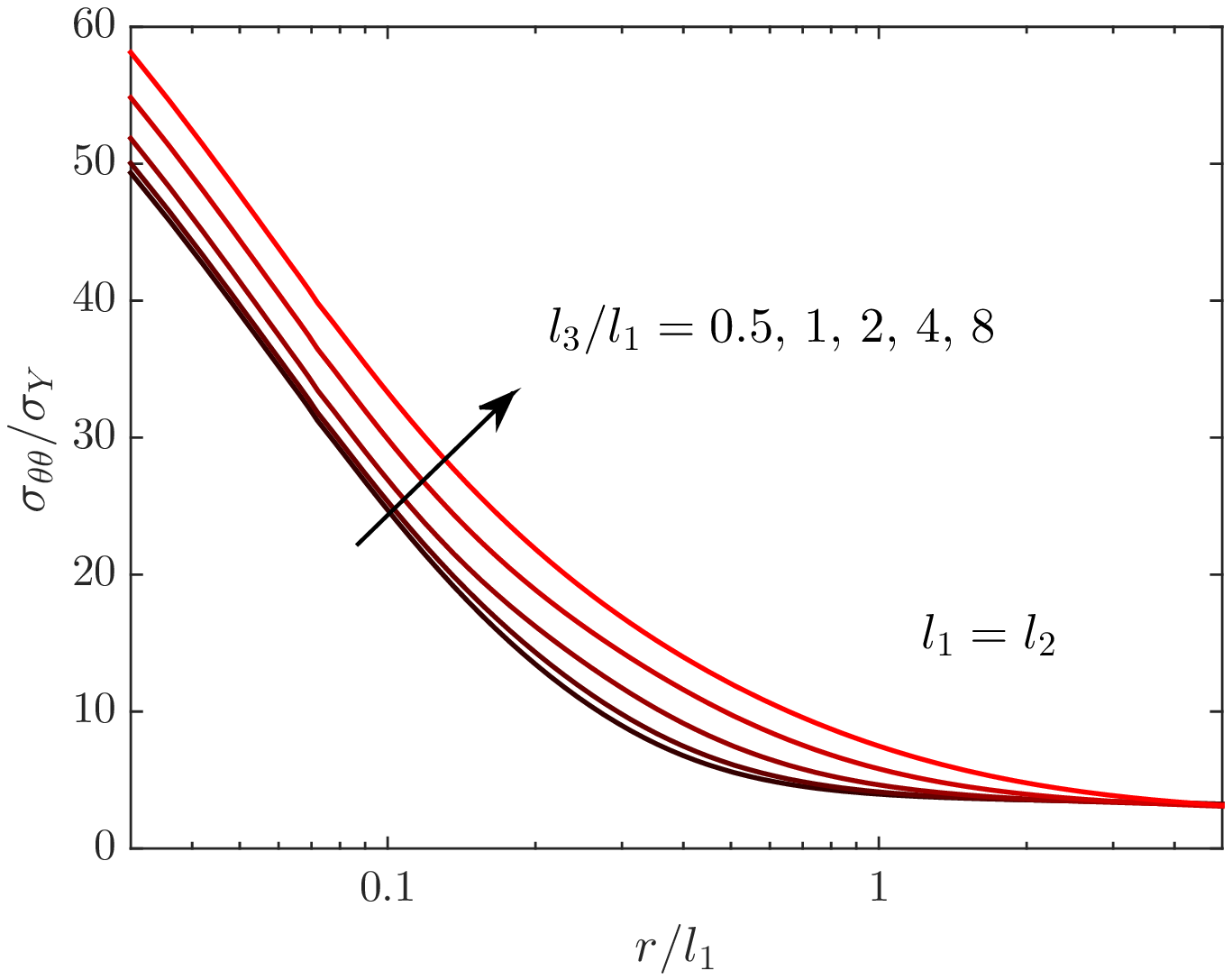}
                \caption{}
                \label{fig:Fig10c}
        \end{subfigure}        
        }
       
        \caption{Opening stress distributions from the phenomenological multiple parameter theory for (a) fixed $l_2$ and $l_3$ ($l_2=l_3$) and varying $l_1$, (b) fixed $l_1$ and $l_3$ ($l_1=l_3$) and varying $l_2$ and (c) fixed $l_1$ and $l_2$ ($l_1=l_2$) and varying $l_3$. For $\sigma_Y=0.2\%E$, $\nu=0.3$, $N=0.2$ and $K_{I}=25\sigma_Y \sqrt{l}$.}\label{fig:Fig10}
\end{figure}

With the aim of gaining insight into the role of individual length scales in the phenomenological three parameter theory, crack tip stress distributions are obtained for various combinations of the length scale parameters. In fig. \ref{fig:Fig10} the influence of each of the parameters is examined by varying its value and keeping fixed the remaining two length scales.\\

\newpage
\newpage
\begin{figure}[H]
\makebox[\linewidth][c]{%
        \begin{subfigure}[b]{0.38\textwidth}
                \centering
                \includegraphics[scale=0.36]{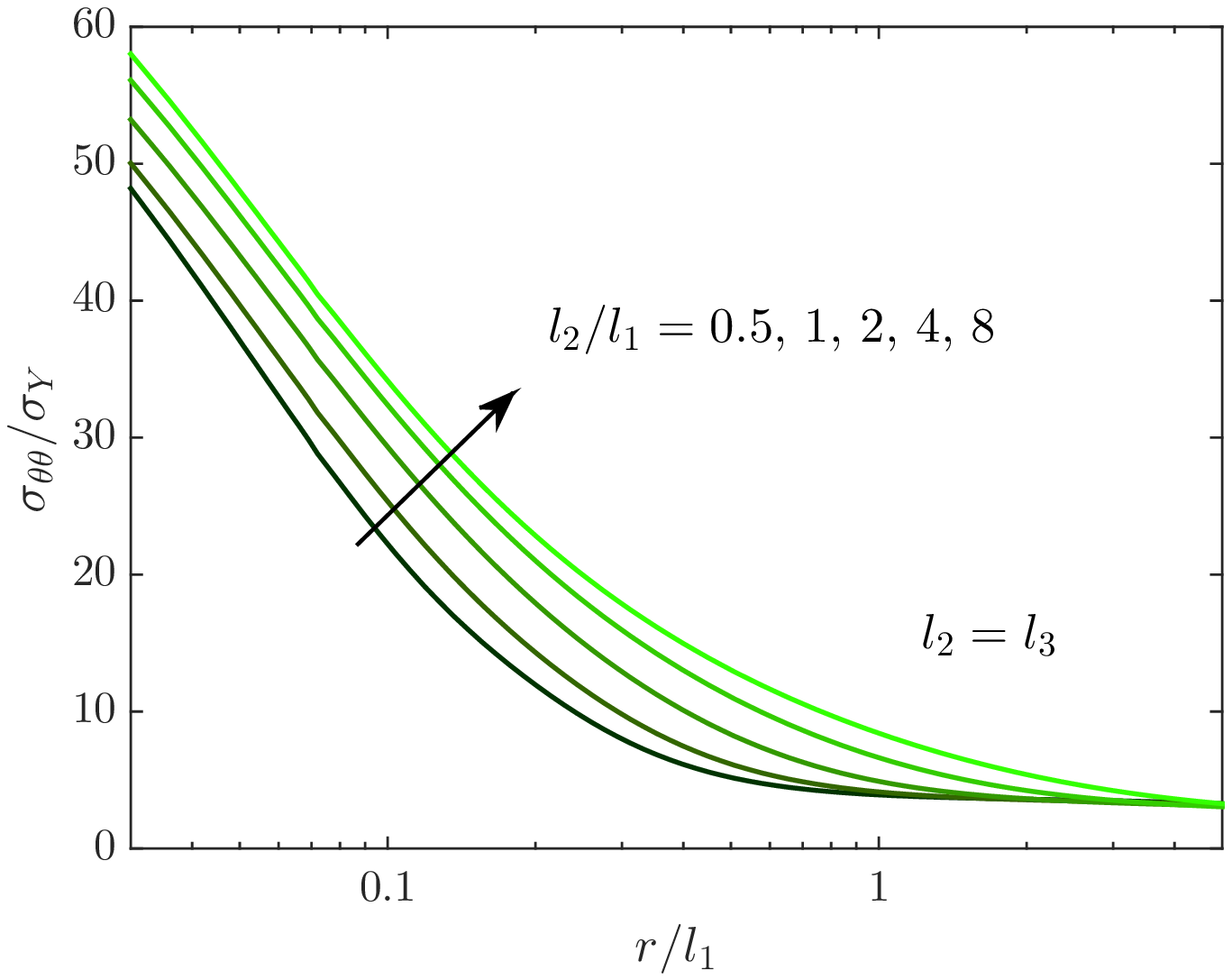}
                \caption{}
                \label{fig:Fig11a}
        \end{subfigure}
        \begin{subfigure}[b]{0.38\textwidth}
                \centering
                \includegraphics[scale=0.36]{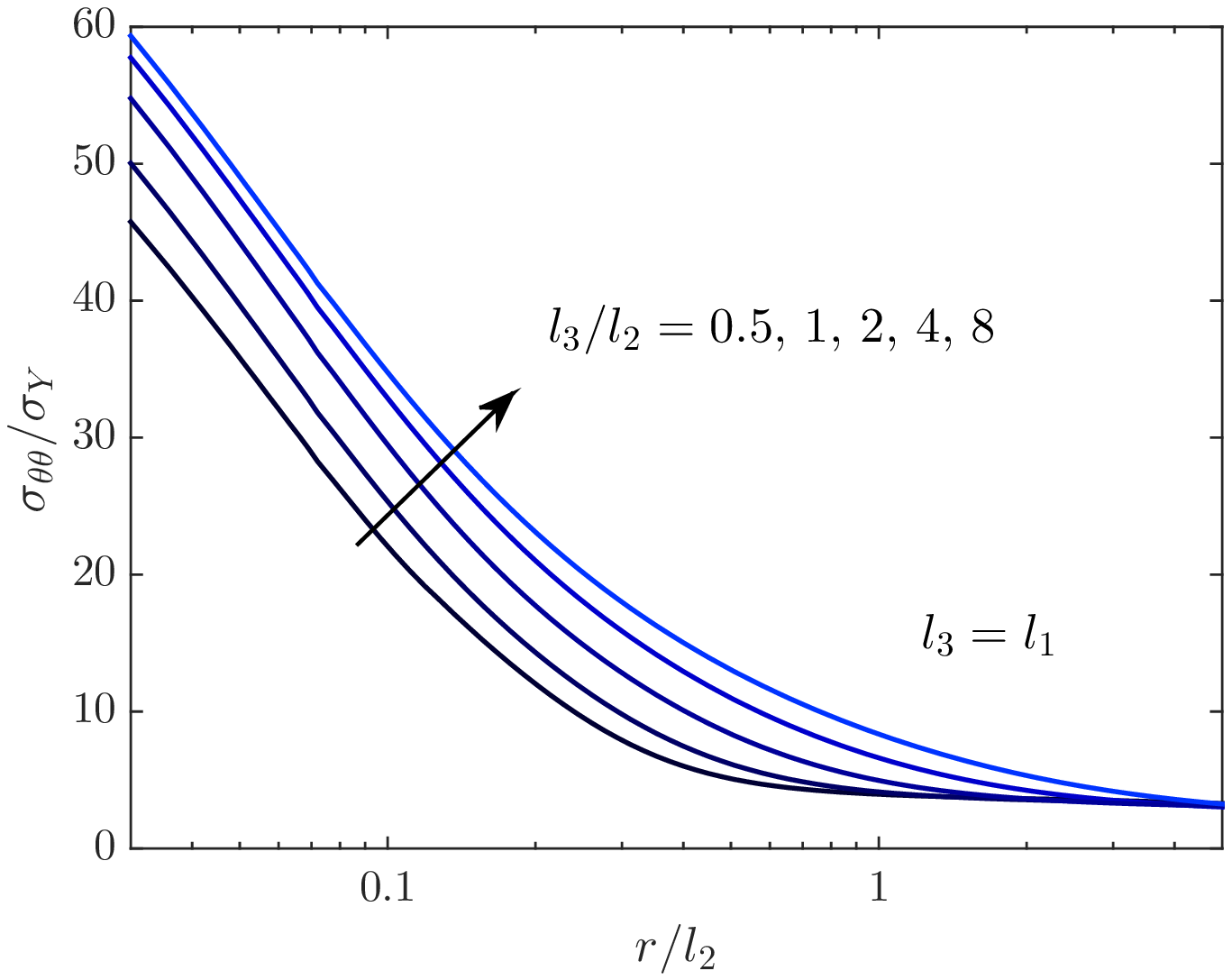}
                \caption{}
                \label{fig:Fig11b}
        \end{subfigure}
        \begin{subfigure}[b]{0.38\textwidth}
                \centering
                \includegraphics[scale=0.36]{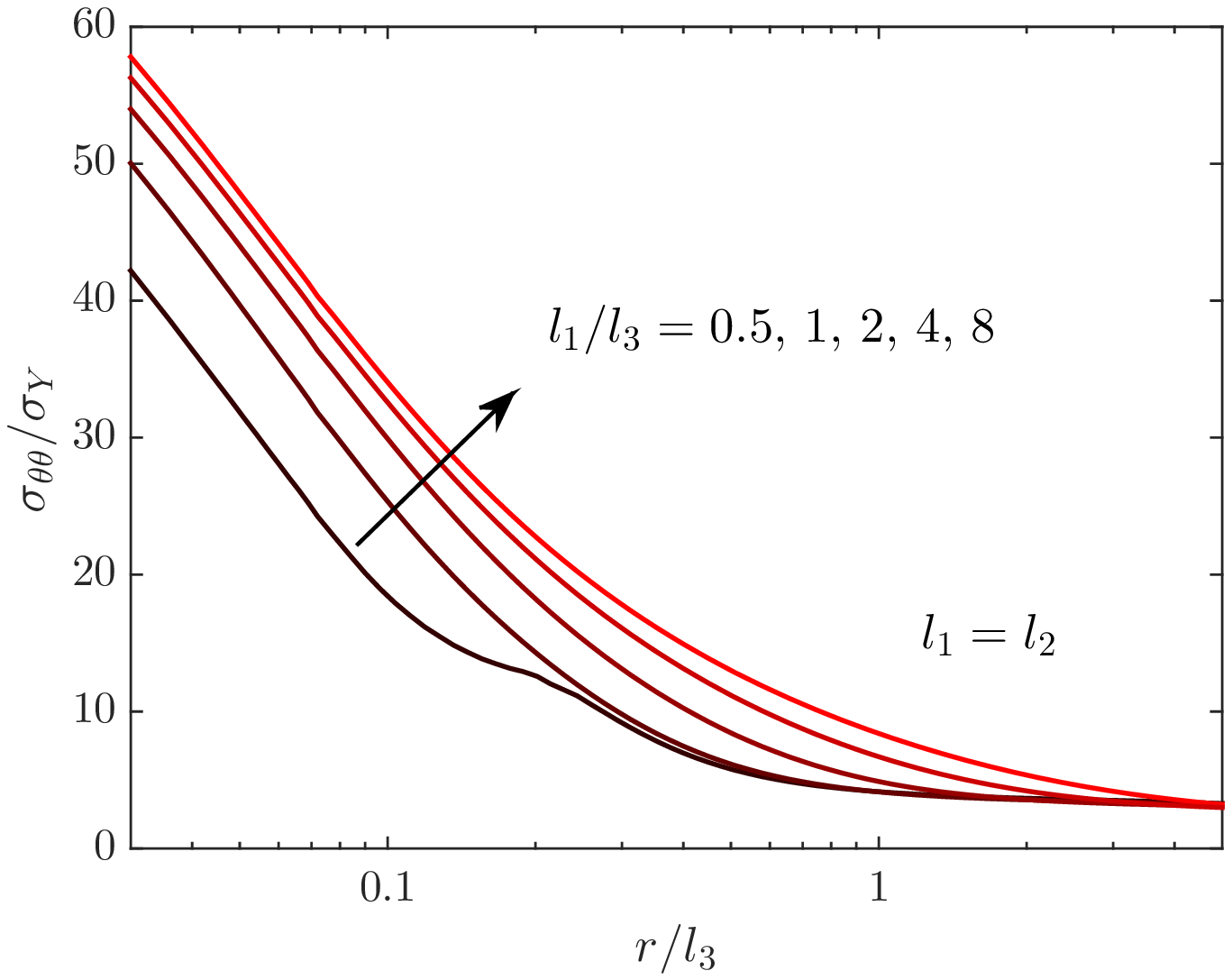}
                \caption{}
                \label{fig:Fig11c}
        \end{subfigure}        
        }
       
        \caption{Opening stress distributions from the phenomenological multiple parameter theory for (a) fixed $l_1$ and varying $l_2$ and  $l_3$ ($l_2=l_3$), (b) fixed $l_2$ and varying $l_1$ and  $l_3$ ($l_1=l_3$), and (c) fixed $l_3$ and varying $l_1$ and  $l_2$ ($l_1=l_2$). For $\sigma_Y=0.2\%E$, $\nu=0.3$, $N=0.2$ and $K_{I}=25\sigma_Y \sqrt{l}$.}\label{fig:Fig11}
\end{figure}
\newpage
From the spread of the curves it is seen that the degree of stress elevation is more sensitive to the first parameter $l_1$ (fig. \ref{fig:Fig10a}), while $l_2$ (fig. \ref{fig:Fig10b}) and $l_3$ (fig. \ref{fig:Fig10c}) play a less relevant role (with $\sigma_{\vartheta \vartheta}/\sigma_Y$ ranging from 46 to 58.5 at, e.g., $r/l=0.03$ versus 49 to 56 and 49.5 to 58, respectively). This behavior may be better appreciated in fig. \ref{fig:Fig11}, where one parameter is fixed and other two parameters are equally varied. Thus, fig. \ref{fig:Fig11a} shows the stress distributions obtained when $l_1$ is kept constant. The comparison with figs. \ref{fig:Fig11b} (constant $l_2$) and \ref{fig:Fig11c} (constant $l_3$) immediately reveals smaller changes in the results when $l_1$ is fixed. Varying $l_2$ or $l_3$ has a similar influence on the results.\\

The major role of $l_1$, the predominant material length in the presence of stretch gradients, supports previous findings by \citet{K08} within the sharp crack problem. This further implies that the combination of length scales that characterizes the influence of strain gradients ahead of the crack must be obtained from indentation testing, where the dominating effect of $l_1$ is also seen (see \citealp{BH98}).\\

Finally, it is necessary to remark that phenomenological higher order modeling of size effects in metal plasticity is under continuous development. While crack tip fields are generally investigated under monotonic and highly proportional loading conditions, one must note that the \citet{FH01} theory was found, under some non-proportional straining histories, to violate the thermodynamic requirement that plastic dissipation must always be non-negative. Positive plastic work was ensured by employing dissipative higher order stresses constitutively related to increments of strain \citep{GU04,G04}. However, it has been very recently noticed that this \textit{non-incremental} formulation may lead to a delay in plastic flow under certain non-proportional loading conditions \citep{F14,BP15}. As the field evolves the role of novel SGP formulations on crack tip mechanics must be assessed. Moreover, the use of single crystal theories (e.g.,\citealp{B06,GR14,WB15}) will certainly provide important insight into the influence of geometrically necessary dislocations in the fracture process zone.

\section{Conclusions}
\label{Concluding remarks}

Large gradients of plastic strain close to the crack tip must undoubtedly lead to additional hardening and very high crack tip stresses that classical plasticity is unable to predict. The experimental observation of cleavage fracture in the presence of significant plastic flow and the experimentally assessed domain where hydrogen cracking nucleates support the concept of an increased dislocation density due to GNDs in the vicinity of the crack.\\

In this work a general framework for damage and fracture assessment including the effect of strain gradients is provided. The numerical scheme of the two main approaches within continuum strain gradient plasticity modeling is developed so as to account for large strains and rotations and differences among theories are revealed and discussed. The following aspects must be highlighted:\\

- Due to the contribution of strain gradients to the work hardening of the material, crack tip blunting is largely reduced and the stress reduction intrinsic to conventional plasticity avoided. This significantly increases the differences with classical plasticity solutions reported in the literature within the infinitesimal deformation framework.\\

- The physical length ahead of the crack where SGP predictions deviate from the estimations of classical plasticity can span several tens of $\mu m$, embracing the critical distance of many damage mechanisms. The magnitude of stress elevation close to the crack tip suggests that accounting for the effect of GNDs in the modelization can be particularly relevant in hydrogen assisted cracking, where damage takes place within 1 $\mu m$ to the crack tip. \\

- Results reveal significant quantitative differences among SGP theories for the same material length scale ($l_1=l_2=l_3=l_{MSG}=l^*$). Within the phenomenological approach, the single length parameter version predicts much smaller size effects than its multiple length parameter counterpart. Estimations from MSG plasticity lead to lower crack tip stresses but a larger gradient dominated zone, relative to the phenomenological predictions. Further research and experimental data are needed to gain insight into the existing correlation between the length scales inferred from each theory.\\

- A dominant effect of the first invariant of the strain gradient tensor is observed in the multiple length parameter version of the phenomenological SGP theory. Since $l_1$ also plays an important role in indentation testing, results indicate that the constitutive length parameters that govern the influence of strain gradients in mode I fracture problems should be inferred from nanoindentation.\\

\section{Acknowledgments}
\label{Acknowledge of funding}

E. Mart\'{\i}nez-Pa\~neda gratefully acknowledges financial support from the Ministry of Science and Innovation of Spain through grant MAT2011-28796-CO3-03, and the University of Oviedo through grant UNOV-13-PF and an excellence mobility grant within the International Campus of Excellence programme. C. F. Niordson gratefully acknowledges financial support from the Danish Council for Independent Research under the research career programme Sapere Aude in the project ``Higher Order Theories in Solid Mechanics".




\end{document}